\documentclass[aps,prl,reprint,amsmath,amssymb,superscriptaddress,noeprint]{revtex4-2}

\usepackage{soul}
\usepackage{graphicx}
\usepackage{dcolumn}
\usepackage{bm}
\usepackage[version=3]{mhchem}
\usepackage{mathtools}
\PassOptionsToPackage{hyphens}{url}
\usepackage[colorlinks=true,allcolors=blue]{hyperref}

\usepackage{lineno}
\usepackage{float}
\usepackage{cleveref}
\usepackage{color}
\usepackage{graphicx}
\usepackage{amsmath,amssymb}
\usepackage{upgreek}
\usepackage{psfrag} 
\usepackage{graphicx}
\usepackage{braket}
\usepackage{units}
\usepackage{times}
\usepackage[svgnames]{xcolor}
\definecolor{myColor}{rgb}{0.02,0.12,0.3}
\definecolor{myciteColor}{rgb}{0.39,0.7,0.89}
\usepackage[colorlinks=true,allcolors=blue]{hyperref}
\newcommand{\F}{\mathrm{F}}
\newcommand{\R}{\mathrm{R}}

\newcommand{\dd }{\mathrm{d}}

\newcommand{\out}{\text{o}}
\newcommand{\kk}{\textbf{k}}

\newcommand{\EF}{E_{\mathrm{F}}}
\newcommand{\as}{a_\mathrm{s}}
\newcommand{\tsf}{\alpha}

\makeatletter
\def\maketitle{
\@author@finish
\title@column\titleblock@produce
\suppressfloats[t]}
\begin{document}

\title{Emergence of Fermi’s Golden Rule in the Probing of a Quantum Many-Body System}
\author{Jianyi Chen}
\email[To whom correspondence should be addressed:\\ derek.chen.jc3757@yale.edu,~songtao.huang@yale.edu.]{}
\affiliation{Department of Physics, Yale University, New Haven, Connecticut 06520, USA}
\author{Songtao Huang}
\email[To whom correspondence should be addressed:\\ derek.chen.jc3757@yale.edu,~songtao.huang@yale.edu.]{}
\affiliation{Department of Physics, Yale University, New Haven, Connecticut 06520, USA}
\author{Yunpeng Ji}
\affiliation{Department of Physics, Yale University, New Haven, Connecticut 06520, USA}
\author{Grant L. Schumacher}
\affiliation{Department of Physics, Yale University, New Haven, Connecticut 06520, USA}
\author{Alan Tsidilkovski}
\affiliation{Department of Physics, Yale University, New Haven, Connecticut 06520, USA}
\author{Alexander Schuckert}
\affiliation{Joint Quantum Institute and Joint Center for Quantum Information and Computer Science, NIST/University of Maryland, College Park, Maryland 20742, USA}
\author{Gabriel G. T. Assumpção}
\affiliation{Department of Physics, Yale University, New Haven, Connecticut 06520, USA}
\author{Nir Navon}
\affiliation{Department of Physics, Yale University, New Haven, Connecticut 06520, USA}
\affiliation{Yale Quantum Institute, Yale University, New Haven, Connecticut 06520, USA}

\date{\today}
\begin{abstract}

Fermi’s Golden Rule (FGR) is one of the most impactful formulas in quantum mechanics, providing a link between easy-to-measure observables – such as transition rates – and fundamental microscopic properties – such as density of states or spectral functions. Its validity relies on three key assumptions: the existence of a continuum, an appropriate time window, and a weak coupling. Understanding the regime of validity of FGR is critical for the proper interpretation of most spectroscopic experiments. While the assumptions underlying FGR are straightforward to analyze in simple models, their applicability is significantly more complex in quantum many-body systems. 
Here, we observe the emergence and breakdown of FGR, using a strongly interacting homogeneous spin-$1/2$ Fermi gas coupled to a radio-frequency (rf) field. Measuring the transition probability into an outcoupled internal state, we map the system's dynamical response diagram versus the rf-pulse duration $t$ and Rabi frequency $\Omega_0$. For weak drives, we identify three regimes: an early-time regime where the transition probability takes off as $t^2$, an intermediate-time FGR regime, and a long-time non-perturbative regime. Beyond a threshold Rabi frequency, Rabi oscillations appear. Our results provide a blueprint for the applicability of linear response theory to the spectroscopy of quantum many-body systems.

\end{abstract}

\maketitle
The framework of linear response theory is central to our understanding of how to interpret probes of complex quantum systems. For instance, connecting experimental observables to many-body quantities often relies on linear response theory via Fermi's Golden Rule (FGR)~\cite{dirac1927quantum,fermi1950nuclear}. FGR states that the long-time effect of a perturbation on a many-body system is given by the Fourier transform of the time correlation function of the perturbation operator with the observable. Depending on the context, this can provide crucial microscopic information on the many-body system in equilibrium, such as spectral functions, dynamic structure factors or local density of states (see \emph{e.g.}~\cite{Pines_Nozières_2018,fetter2012quantum,bruus2004many}). 
The applicability of FGR is wide, ranging from describing scanning tunneling microscopy~\cite{STM_1, STM_2}, optical conductivity measurements~\cite{jankovic2014nonequilibrium}, to inelastic scattering reactions in nuclear physics~\cite{nuclear_physics_FGR}. 

The regime of validity of FGR is a problem of fundamental interest in quantum many-body physics; in strongly correlated systems, determining the validity of FGR \emph{a priori}\ is a formidable challenge. 
FGR generically provides a transition rate from an initial state into a continuum under a coupling perturbation. 
The three underlying assumptions - a continuum to decay to, a time neither too long nor too short, and a weak coupling — are crucial. Several scenarios where these assumptions are violated have attracted theoretical interest. 
First, the applicability of FGR to systems with a discrete quasi-continuum was recently studied, \emph{e.g.}\ in the context of thermalization of finite-size isolated many-body systems (see \emph{e.g.}~\cite{heating_Rigol,rigol2021,Micklitz_2022}). 
Secondly, at short times, it is known that FGR does not apply due to unitary time evolution~\cite{debierre2021,aspect_2010}. At long times (\emph{i.e.}\ compared to the inverse level spacing of the quasi-continuum), FGR also breaks down - even if first-order perturbation theory might still apply~\cite{zhang2016}. Thirdly, in some situations, effective couplings might never be weak enough for FGR to apply, even in weakly interacting systems (see \emph{e.g.}\ phonons in a weakly interacting 2D superfluid ~\cite{castin2023}).

Despite FGR's pivotal role, there is a scarcity of experimental studies of its applicability. 
Ultracold gases are attractive platforms to investigate such problems since their Hamiltonians are well known and coupling terms can be accurately engineered~\cite{vale_2021spectroscopic}. Furthermore, fast timescales associated with out-of-equilibrium many-body dynamics are often directly accessible, see \emph{e.g.}~\cite{makotyn2014,Cetina_2016,eigen2018,dyke2021,Vivanco_2023}. 
Here, we map out the full dynamical response diagram 
of a strongly correlated quantum many-body system - a universal spin-1/2 Fermi gas - to a time-dependent momentum-independent coupling, and observe the emergence and breakdown of FGR (Fig.~\ref{Fig1}).

\begin{figure}[!hbt]
\includegraphics[width=1\columnwidth]{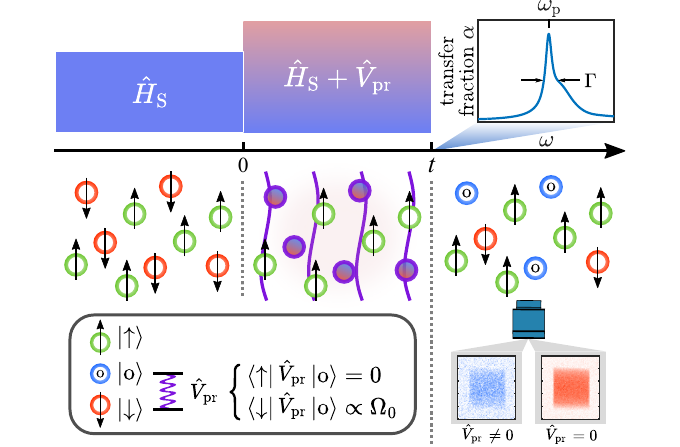}
\caption{Probing a quantum many-body system: the system considered here is a homogeneous universal spin-$1/2$ Fermi gas, prepared in equilibrium of the Hamiltonian $\hat{H}_\text{S}$, in the two internal states $\ket{\uparrow}$ and $\ket{\downarrow}$ (green and red circles). A probe $\hat{V}_{\mathrm{pr}}$ is turned on for a duration $t$, which couples $\ket{\downarrow}$ to the initially unoccupied state $\ket{\out}$ (blue circles), with a (momentum-independent) bare-transition Rabi frequency $\Omega_0$ and a detuning $\omega$ from the bare transition. The transfer fraction $\tsf$ is then measured, see typical absorption images of $^6$Li atoms in a box trap in the bottom right (the color is associated to the corresponding spin state, encoded in different Zeeman sublevels). A cartoon spectrum is shown in the top right, characterized by a peak frequency shift $\omega_\mathrm{p}$ and width $\Gamma$. 
}
\label{Fig1}
\end{figure}

\begin{figure}[!hbt]
\includegraphics[width=1\columnwidth]{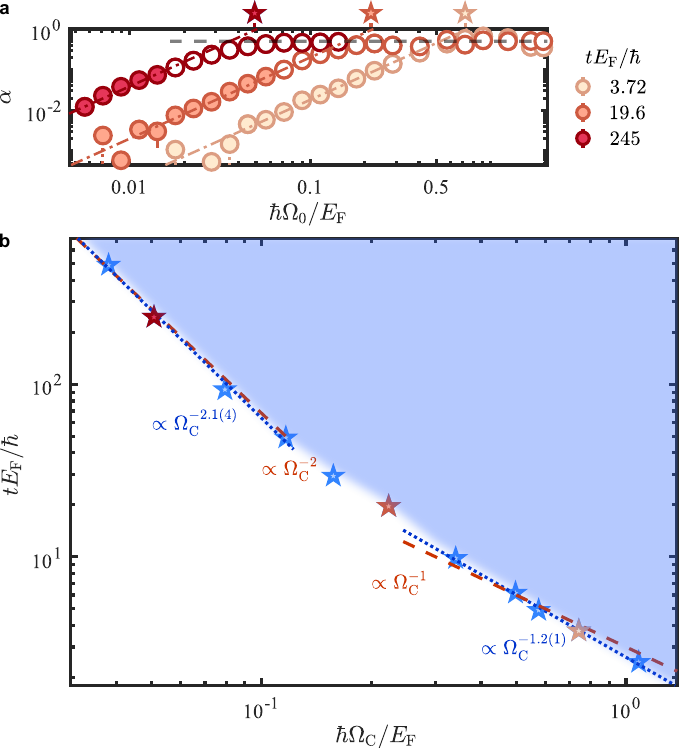}
\caption{Boundary between the linear and nonlinear response regimes. (\textbf{a}) Transfer fraction $\tsf$ versus $\hbar\Omega_0/\EF$ for various $t\EF/\hbar$. The dash-dotted lines are fits $(\Omega_0/\Omega_\mathrm{C})^2$ to the data points restricted to $\tsf<0.1$ (solid markers). The gray dashed line is the large-$\Omega_0$ limit $\tsf = 1/2$. The dotted error bars in the bottom left are for points for which the transfer fraction is not statistically distinguished from $0$. (\textbf{b}) Boundary $\Omega_\mathrm{C}$ versus $t$. The nonlinear (resp. linear) response regime is marked as the blue-shaded (resp. white) area. The blue dotted lines are power-law fits in the low-$\Omega_\mathrm{C}$ and high-$\Omega_\mathrm{C}$ limits; the extracted exponents are shown in adjacent legends. The red dashed lines are the theoretical predictions, see text.
The red, orange, and light orange points correspond to the data sets in (\textbf{a}).}
\label{Fig2}
\end{figure}

Traditionally, FGR arises from a linear response analysis on a system governed by a Hamiltonian $\hat{H}_{\mathrm{S}}$ perturbed by a probing term $\hat{V}_{\mathrm{pr}}$. We assume that $\hat{V}_{\mathrm{pr}}$ is turned on at $t=0$ and outcouples particles in spin state $\ket{\downarrow}$ into a noninteracting state $\ket{\out}$ with driving frequency $\omega_{\mathrm{pr}}$ as $\hat{V}_{\mathrm{pr}}=  \hbar\Omega_0  \cos\left(\omega_\mathrm{pr} t\right)\sum_{\textbf{k}} \hat{c}_{\text{o},\textbf{k}}^\dagger \hat{c}_{\downarrow,\textbf{k}}+\mathrm{h.c.}$, where $\Omega_0$ is the bare-transition Rabi frequency characterizing the coupling strength and $\hat{c}_{\sigma,\textbf{k}}$ (resp. $\hat{c}^\dagger_{\sigma,\textbf{k}})$ is the annihilation (resp. creation) operator for particles with spin state $\sigma$ and momentum $\hbar\kk$; note that $\hat{V}_{\mathrm{pr}}$ does not couple to spin state $\ket{\uparrow}$. 
The main observable for studying the dynamics induced by $\hat{V}_{\mathrm{pr}}$ is the transition rate $R(\omega,t)\equiv \mathrm{d\alpha}/\mathrm{d}t$ of the fraction of outcoupled particles $\tsf\equiv N_{\out}(\omega,t)/N_\downarrow$, where $N_\downarrow$ is the initial number of particles in state $\ket{\downarrow}$, $t$ is the coupling pulse duration, and $\omega=\omega_\mathrm{pr}-\omega_0$ is the detuning to the bare internal-state transition. 
In a suitable time window,
FGR provides a powerful expression for (a time-independent) $R$~\cite{Haussmann_2009,Törmä_2016}:
\begin{equation}
R(\omega,t)\rightarrow
R_\mathrm{FGR}(\omega)=\frac{\Omega_0^2}{4N_{\downarrow}}\sum_\textbf{k}A(\textbf{k},\xi_{k\downarrow}/\hbar-\omega)n_{\mathrm{F}}(\xi_{k\downarrow}-\hbar\omega),
\label{Irf_FGR}
\end{equation}
where $\xi_{k\downarrow} = \hbar^2 k^2/(2m) - \mu_{\downarrow}$, $k = \lvert \textbf{k}\rvert$, $m$ is the atom's mass, $\mu_{\downarrow}$ is the chemical potential, and $n_{\mathrm{F}}(\epsilon)$ is the Fermi-Dirac distribution at energy $\epsilon$. The most interesting quantity in Eq.~\eqref{Irf_FGR} is the spectral function $A(\textbf{k},\omega')$, which encodes information on the excitations and the density of states of a many-body system~\cite{fetter2012quantum,SuppMat}.  

Our platform for exploring this problem is a homogeneous unitary Fermi gas in a spin-population balanced mixture of $\ket{\downarrow}$ and $\ket{\uparrow}$. In the energy diagram of $^6$Li, the internal states $\{\ket{\downarrow},\ket{\out},\ket{\uparrow}\}$ correspond respectively to the three lowest Zeeman sublevels. 
We work at a bias magnetic field $B \approx 690$~G, for which the s-wave scattering length $\as$ between $\ket{\downarrow}$ and $\ket{\uparrow}$ is infinite, $1/\as=0$~\cite{zwerger_2011}. We typically have spin populations of $N_{\downarrow} \approx N_{\uparrow} \approx 2\times 10^5$ atoms, and a temperature $T \approx 0.15E_{\mathrm{F}}/k_{\mathrm{B}}$, where the Fermi energy is $E_{\mathrm{F}} \approx h\times 4~$kHz. 
The coupling $\hat{V}_\mathrm{pr}$ is realized with a radio-frequency (rf) field~\cite{Note1}; we verified that the trapping is uniform both for the initial spin states and for $\ket{\out}$~\cite{SuppMat}. 
We measure $\tsf$ as a function of the control parameters $\Omega_0$ and $t$, fixing the rf detuning to $\omega=0.63(2) E_{\mathrm{F}}/\hbar$; this value corresponds to the peak frequency shift in the weak-drive limit, \emph{i.e.}\ $\omega_\mathrm{p}$ for small $\Omega_0$ (see cartoon of the spectrum in Fig.~\ref{Fig1}~\cite{Note2}). 

We first map the boundary where the linear response framework breaks down. Since perturbation theory generically predicts a linear relationship of $R$ with driving power, the key signature of the nonlinear response regime is that the transfer fraction is $\tsf \not\propto \Omega_0^2$. In Fig.~\ref{Fig2}a, we show $\tsf$ as a function of $\Omega_0$ for a few selected $t$; here and wherever relevant, we normalize quantities relative to $E_\F$, the natural energy scale of this system. While at large $\Omega_0$, all transfer fractions approach the asymptotic value of $1/2$ (see the gray dashed line), for small enough $\Omega_0$, all series feature a clear region where $\tsf \propto \Omega_0^2$. We determine this (smooth) boundary by fitting the small-$\tsf$ data with $\tsf=(\Omega_0/\Omega_\mathrm{C})^2$. The adjustable parameter $\Omega_\mathrm{C}$ is a convenient parameter that characterizes the crossover from the perturbative (linear response) to the non-perturbative (nonlinear response) regimes. The extracted $\Omega_\mathrm{C}$ for various pulse durations is shown in Fig.~\ref{Fig2}b. Notably, we observe power laws for both weak and strong drives. At strong drives, a power-law fit gives $t\propto\Omega_\mathrm{C}^{-1.2(1)}$; this agrees with the expectation that $\Omega_\mathrm{C}t \sim 1$ in the large-$\Omega_0$ regime where the quantum dynamics would be dominated by Rabi oscillations between the internal states $\ket{\downarrow}$-$\ket{\out}$. For weak drives, we find instead $t \propto \Omega_{\mathrm{C}}^{-2.1(4)}$, which agrees with the expectation that $\alpha\propto \Omega_0^2 t$ in FGR's regime.

\begin{figure}[!hbt]
\includegraphics[width=1\columnwidth]{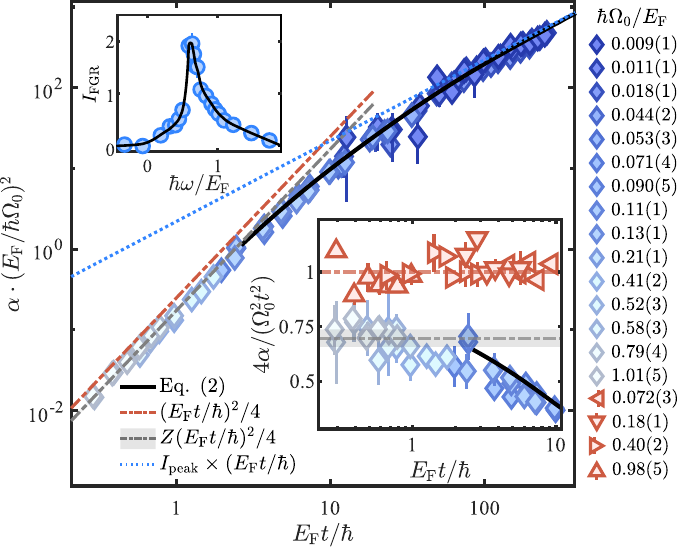}
\caption{Dynamics in the linear response regime and emergence of FGR. Time series of the transfer fraction $\alpha(t)$ are shown for various $\Omega_0$ and postselected such that $\alpha<0.1$ and $t<12~\mathrm{ms}$~\cite{SuppMat}. Both in the main panel and the lower inset,
the red (resp. gray) dash-dotted line is the universal quadratic $\tsf = \frac{1}{4} \Omega_0^2 t^2$ at early times (resp. a fit $\tsf = \frac{Z}{4} \Omega_0^2 t^2$, where $Z=0.70(4)$ and the band represents the uncertainty). Upper inset: spectral response $I_\text{FGR}(\omega)$ measured in FGR's regime (see text). Near the peak, we use $\Omega_0 = 2\pi\times45(3)~\mathrm{Hz}$ for $t = 5~\mathrm{ms}$. Away from the peak, we typically use $\Omega_0 = 2\pi\times 140(5)~\mathrm{Hz}$, for $t = 1.5~\mathrm{ms}$ to improve the signal~\cite{SuppMat}. The solid line is an interpolation used to determine $\alpha$ via Eq.~(\ref{relation_N_I}); see the corresponding solid black lines in the main panel and lower inset. As the short-time behavior of $\tsf$ is dictated by the high-$\omega$ part of the rf spectrum, the solid black lines are limited by the range of $\omega$ over which $I_{\mathrm{FGR}}$ was measured (see~\cite{SuppMat} for a discussion of extrapolations). The dotted blue line in the main panel is determined from the peak spectral response $I_\text{peak}$. Lower inset: Early-time dynamics. The ideal-gas data (in practice, a single-component Fermi gas) is shown as red points and a fit gives $\tsf/\Omega_0^2 = 0.26(2) \times t^{2.02(5)}$, in excellent agreement with the theoretical limit. 
}
\label{Fig3}
\end{figure}

We now use this boundary to focus on the time dynamics within the linear response regime. In Fig.~\ref{Fig3}, we show the transfer fraction $\tsf\cdot(\EF/\hbar\Omega_0)^2$ - the normalization is chosen for the perturbative regime -
as a function of the normalized time $\EF t/\hbar$ for various $\Omega_0$. We only display data such that $\tsf < 0.1$ and observe that all data sets collapse onto a unique curve. 

At long times ($E_{\mathrm{F}}t/\hbar \gtrsim 70$), the normalized $\tsf$ approaches a scaling $\propto t$ characteristic of FGR's regime (see the dotted blue line in Fig.~\ref{Fig3}), irrespective of $\Omega_0$. In the regime where $\tsf\propto \Omega_0^2 t$, we measure the dimensionless linear response spectrum 
$I_\text{FGR} \equiv \tsf \cdot (E_{\mathrm{F}}/\hbar\Omega_0^2t)=R_\mathrm{FGR}(\omega)\cdot (E_{\mathrm{F}}/\hbar\Omega_0^2)$ as a function of the normalized detuning $\hbar\omega/E_{\mathrm{F}}$, see upper inset of Fig.~\ref{Fig3}~\cite{Note3}.

At short times ($E_{\mathrm{F}}t/\hbar \lesssim 3$), the transfer fraction exhibits a $t^2$ scaling. 
At extremely short times, we expect the universal limit $\tsf\cdot(\EF/\hbar\Omega_0)^2 \approx (E_{\mathrm{F}}t/\hbar)^2/4$, which reflects the normalization of the spectral function~\cite{SuppMat} (red dot-dashed line). In the lower inset, we show the data taken for the ideal gas (red points), which closely follows that expectation (red dot-dashed line). However, for the shortest times accessed in our work, we find that the interacting-gas measurements are instead well described by $Z \cdot(E_{\mathrm{F}}t/\hbar)^2/4$, with $Z=0.70(4)$ (gray dot-dashed line and band). A well-defined value of $Z$ at intermediate times is indicative of a separation of scales in the spectral response, with a well-defined peak of area $Z$~\cite{Note4}. 

More generally, linear response theory implies a powerful link between FGR's spectrum of Eq.~(\ref{Irf_FGR}) - presumably valid only at long times - and the dynamics of the transfer fraction $\tsf(\omega,t)$ at all times $t$~\cite{SuppMat}:
\begin{equation}
\begin{split}
    \tsf(\omega,t)
    =& ~t^2\int_{-\infty}^\infty\frac{\dd\omega'}{2\pi}~\mathrm{sinc}^2\left(\frac{\omega'-\omega}{2}t\right)R_\mathrm{FGR}(\omega'),
\end{split}
\label{relation_N_I}
\end{equation}
where $\mathrm{sinc}(x)\equiv \sin(x)/x$.  
This result is the many-body counterpart of the textbook finite-time form of FGR~\cite{cohen1998atom}. 
Based on a generic form for $R_\mathrm{FGR}$ of typical width $\Gamma$ (inset of Fig.~\ref{Fig3} and cartoon of Fig.~\ref{Fig1}), we can qualitatively discuss the various spectroscopic regimes, at a fixed detuning $\omega$. In this work, we focus on the near-resonant case $\omega\approx\omega_\mathrm{p}$.

When $t\gg \Gamma^{-1}$, the $\mathrm{sinc}^2$ becomes much narrower than $R_\mathrm{FGR}$. This is \emph{FGR's regime}, for which Eq.~(\ref{relation_N_I}) implies that the transfer fraction $\tsf=\lambda t$, where $\lambda\equiv R_\mathrm{FGR}\propto \Omega_0^2$ is FGR's transition rate. Note that $\Omega_0$ has to remain low enough for Eq.~(\ref{relation_N_I}) to be valid (especially $\tsf\ll 1$ so that $\lambda t\ll 1$). Secondly, in the limit $t\ll \Gamma^{-1}$, one finds $\tsf = \Omega_0^2t^2/4$, the \emph{(time-)quadratic regime}, which is the early-time behavior expected for a driven two-level system; 
in the strong drive limit, the criterion of small transfer fraction is $\Omega_0t \ll 1$. Both these regimes, characterized by the scaling $\tsf\propto \Omega_0^2$, are described by linear response theory.

Using the experimentally measured $R_\mathrm{FGR}$ (the solid black line interpolation in the upper inset of Fig.~\ref{Fig3}), we can compute $\tsf(\omega,t)$ via Eq.~(\ref{relation_N_I}). The result, shown as the solid black line in the main panel agrees very well with the data, even away from FGR's regime $E_{\mathrm{F}}t/\hbar \lesssim 70$. 

The applicability of FGR has important implications for the interpretation of rf spectra measurements~\cite{Mukherjee_2019,ji_2023_QJT} (see also~\cite{li2024} for the measurement of the same spectrum using a microwave transition). Therefore, we now focus on a more careful examination of the boundary of FGR, in the low-$\Omega_0$ regime. We take spectra analogous to the upper inset of Fig.~\ref{Fig3} for various $\Omega_0$, with a pulse duration $t \approx 2\pi/(5\Omega_0)$; since we remain close to FGR's regime, we also normalize the more general spectra to $I \equiv \tsf\cdot (\EF/\hbar\Omega_0^2t)$. From each of those spectra, we extract the peak frequency shift $\omega_\mathrm{p}$, the (full width at half-maximum) width $\Gamma_{\mathrm{fit}}$ and peak spectral response $I_\mathrm{peak}$ using a smooth interpolation of the data~\cite{SuppMat}; we show the results in Fig.~\ref{Fig4}. Two examples of spectra are shown in the inset of Fig.~\ref{Fig4}, one within (light red points) and one beyond (dark red points) FGR's regime for $\omega\approx\omega_\mathrm{p}$.

The measured $\omega_\mathrm{p}$ is largely insensitive to $\hbar\Omega_0/\EF$;
this justifies \textit{a posteriori} our choice of fixing the rf detuning to the low-$\Omega_0$ value of $\omega_\mathrm{p}$. 
On the other hand, $\Gamma_\text{fit}$ and $I_\text{peak}$ vary significantly, even down to surprisingly low $\hbar\Omega_0/E_\F$. It is only for $\hbar\Omega_0/E_\F\lesssim 10^{-2}$ that $\Gamma_\text{fit}$ and $I_\text{peak}$ level off and one reaches the $\Omega_0$-independent regime.
The gray lines correspond to the same spectral properties calculated from Eq.~\eqref{relation_N_I} (see caption). The good agreement indicates that the variation of the spectral properties even in the linear response regime can be understood by a finite pulse area effect. In particular, we find in the low-power limit that the spectral width is $\hbar\Gamma_0/E_{\mathrm{F}} = 0.24(6)$. Under typical weak drives used before~\cite{Mukherjee_2019,ji_2023_QJT}, $\Gamma_\mathrm{fit}$ can be $50\%$ larger - or more - than $\Gamma_0$. 
Furthermore, we verify the scaling in time by extracting $I_\text{linfit}$, defined as a linear fit to $\tsf(\omega_{\mathrm{p}},t)\cdot \frac{\EF}{\hbar\Omega_0^2}$ up to a pulse duration $t = 2\pi/(5\Omega_0)$, see light orange points in Fig.~\ref{Fig4}. The agreement with $I_\mathrm{peak}$ is very good, which highlights that a clean linear-response-like scaling $\alpha\propto t$ does not guarantee the validity of FGR. This shows that even with seemingly reasonable choices of Rabi frequencies and pulse areas, systematic errors can arise from deviations to FGR. 
\begin{figure}[!hb]
\includegraphics[width=1\columnwidth]{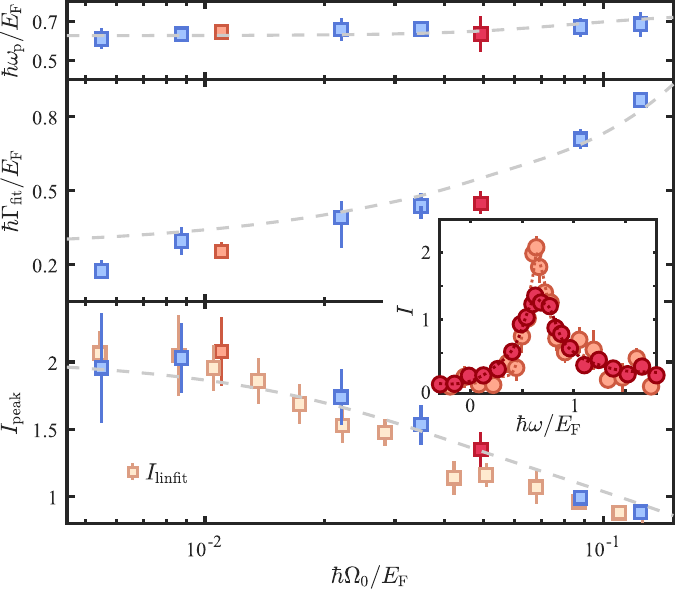}
\caption{Spectral properties near FGR's regime. The normalized peak frequency shift $\hbar\omega_{\mathrm{p}}/\EF$, spectral width $\hbar\Gamma_{\mathrm{fit}}/\EF$, and peak spectral response $I_{\mathrm{peak}}$ are determined from the spectra $I(\omega)$ for various $\Omega_0$ (for a pulse area $\Omega_0 t\approx 2\pi/5$); see examples in the inset. The transfer rate $I_{\mathrm{linfit}}$ is determined from a linear fit to the time series $\tsf(\omega_{\mathrm{p}},t)\cdot \frac{\EF}{\hbar\Omega_0^2}$. The dashed gray lines are calculated from Eq.~(\ref{relation_N_I}) using as input FGR's spectrum of the upper inset of Fig.~\ref{Fig3}, and $t=2\pi/(5\Omega_0)$.
Inset: Examples of spectra within (light red) and beyond (dark red) FGR's regime. Dotted lines show the smooth interpolation~\cite{SuppMat} used for extracting spectral properties. The corresponding $\Omega_0$ of the spectrum can be read from the matching marker color in the main plot.}
\label{Fig4}
\end{figure}

\begin{figure}[!hbt]
\includegraphics[width=1\columnwidth]{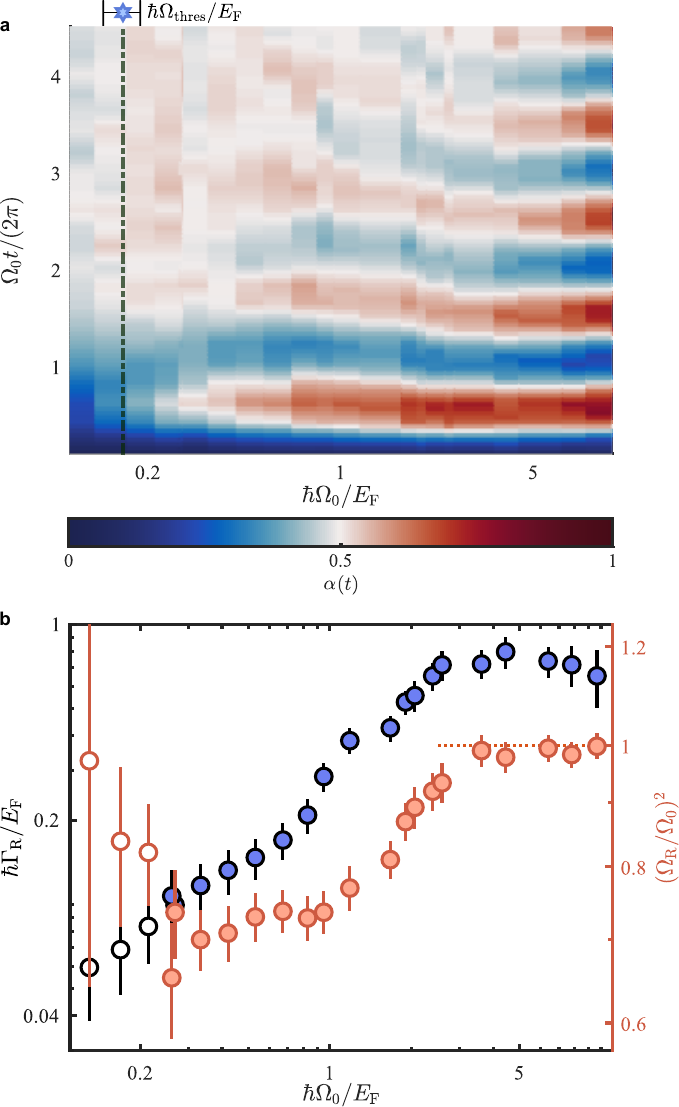}
\caption{Threshold for Rabi oscillations. (\textbf{a}) Time series $\tsf(t)$ for various $\hbar\Omega_0/E_{\mathrm{F}}$. The dash-dotted line and the six-point star indicate the threshold Rabi frequency $\hbar\Omega_{\text{thres}}/E_{\mathrm{F}}$. (\textbf{b}) Oscillation frequency $\Omega_\mathrm{R}$ (light red points) and decay rate $\Gamma_\mathrm{R}$ (blue points) extracted from fits to the time series (see text and~\cite{SuppMat} for details). The open symbols correspond to $\Omega_0\lesssim \Omega_\text{thres}$.}
\label{Fig5}
\end{figure}

\begin{figure}[!hbt]
\includegraphics[width=1\columnwidth]{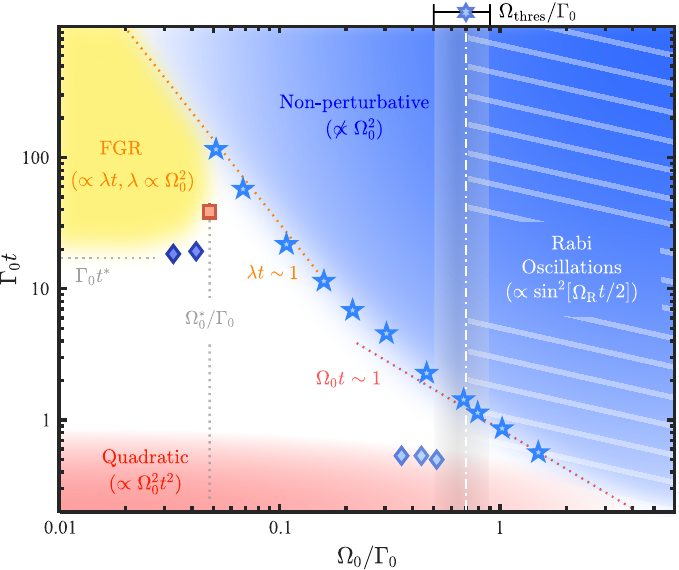}
\caption{Dynamical response diagram of the universal spin-$1/2$ Fermi gas versus the coupling control parameters $\Omega_0$ and $t$. The dotted lines mark the (non-sharp) boundaries between regimes; $\lambda$ and $\Gamma_0$ are FGR's transition rate and spectral width.  
The stars represent the rescaled data from Fig.~\ref{Fig2}, illustrating the condition where $\tsf = 10\%$. Diamond and square markers, which correspond to various measurements from Figs.~\ref{Fig3}-\ref{Fig4}, with matching colors, highlight the boundaries between different regimes. The dash-dotted line and six-point star indicate the threshold frequency for Rabi oscillations as in Fig.~\ref{Fig5}, and the band represents the uncertainty.
}
\label{Fig6}
\end{figure}

For large enough drives, the non-perturbative dynamics will ultimately
consist of Rabi oscillations, which can exhibit many-body effects~\cite{kohstall_2012, Knap_2013, Vivanco_2023}. 
We now examine the emergence of those oscillations. In Fig.~\ref{Fig5}a we show a map of the dynamics of $\tsf(t)$ as a function of $t$ and $\Omega_0$. At low $\Omega_0$, $\tsf(t)$ is mostly monotonically increasing. For $\Omega_0$ larger than a threshold $\Omega_\text{thres}$, Rabi oscillations appear, with an oscillation frequency $\Omega_\text{R}\sim \Omega_0$; we determine that $\hbar\Omega_\text{thres}/\EF= 0.17(2)$ (see the time series data in~\cite{SuppMat}). Calculating this threshold theoretically is a complex out-of-equilibrium many-body problem. However, we remark that $\Omega_\text{thres}/\Gamma_0= 0.7(2)$, which suggests, somewhat analogously to the atom-photon case~\cite{cohen1998atom}, that this transition occurs in a many-body system when the Rabi frequency is comparable to the continuum width. Recent measurements on the highly spin-imbalanced Fermi gas~\cite{Vivanco_2023} (\emph{i.e.}\ $N_\downarrow\ll N_\uparrow$) also support this idea; indeed, from the data in~\cite{Vivanco_2023}, we determine that $\Omega_\text{thres}/\Gamma_0 = 0.8(2)$.

For $\Omega_0>\Omega_\text{thres}$, we extract the oscillation frequency $\Omega_\R$ and the decay rate $\Gamma_\R$ from a fit, see Fig.~\ref{Fig5}b. The renormalized oscillation frequency $(\Omega_\R/\Omega_0)^2$ shows a plateau at low $\Omega_0$, down to $\Omega_0\approx\Omega_\text{thres}$. Its plateau value $(\Omega_\R/\Omega_0)^2 = 0.73(5)$ is close to an analogous measurement in the highly-imbalanced Fermi gas, $0.63(3)$~\cite{Vivanco_2023,scazza2017}. For $\hbar\Omega_0\gtrsim E_\F$, $\Omega_\R/\Omega_0$ rises rapidly. At high $\Omega_0$, $\Omega_\R$ approaches $\Omega_0$; in that regime, $\Omega_0/\Gamma_0 \gtrsim 15$, so that the many-body system can be approximated as a two-level system~\cite{Baym_2007,cohen1998atom}. At low $\Omega_0$, $\Gamma_\R$ decreases with $\Omega_0$, without reaching a nonzero limit. At $\Omega_0\approx 4\EF/\hbar$, $\Gamma_\R$ approaches a maximum, $\hbar\Gamma_\R\approx 0.8\EF$. The analogous measurements in the highly-imbalanced system were connected to many-body properties of the dressed (Fermi polaron) impurity, \emph{e.g.}\ the quasiparticle residue and its lifetime~\cite{Vivanco_2023}. Intriguingly, even though such simple interpretations are elusive in this spin-balanced system (as there is no small parameter, and no well-defined quasiparticles in the normal phase), the results are qualitatively similar~\cite{SuppMat,Note5}.

Finally, we gather all the above results to sketch the dynamical response diagram versus the control parameters $t$ and $\Omega_0$, for $\omega\approx \omega_\mathrm{p}$, see Fig.~\ref{Fig6}~\cite{Note6}. The stars mark the boundary between the linear and non-perturbative (nonlinear) response regimes. The diamonds mark on the low-$t$ end the border of the quadratic regime $\alpha\propto t^2$ (red area) and on the high-$t$ end the border to FGR's regime (golden area). The typical timescale that separates the two is $1/\Gamma_0$, although the prefactors are quite large, \emph{e.g.}\ the onset time for FGR's regime is $t^*\approx 20 \Gamma_0^{-1}$. The square marks $\Omega_0^*$, the largest $\Omega_0$ for which an FGR regime exists; we find $\Omega_0^*\approx 0.05\Gamma_0$. Finally, the six-pointed star indicates the location of $\Omega_\text{thres}$, the threshold for Rabi oscillations (white hatched region)
~\cite{Note7}.

In summary, we mapped out the various dynamical regimes of a strongly correlated quantum many-body system coupled to a probe, revealing the emergence of FGR's regime. The surprisingly low driving power (in natural units) that marks the boundary of FGR's regime serves as a cautionary tale: it shows the importance of careful examination of scalings with time and coupling strength in order to apply FGR's interpretation for the spectroscopy of a complex system. 

The mapping achieved in this work is generic, as it applies to a system with a spectrum consisting of a peak (of spectral weight $<1$) of width $\Gamma$. In the future, it would be interesting to explore the analogous diagram for a system with a more complex spectrum, \emph{e.g.}\ multiple peaks (for instance, on the BEC side, $a_{\mathrm{s}}>0$). Furthermore, exotic early time behaviors, such as two distinct quadratic regimes, could be observed if the spectral response exhibits a scale separation. 

Another interesting avenue is the weak-drive long-time dynamics of this system. While the simpler case of discrete level-to-continuum is amenable to the non-perturbative Wigner-Weisskopf approach~\cite{cohen1998atom,Seke_1988}, our many-body setting is more complex as the long-time spin populations generally tend to a nontrivial mixture of $\{\ket{\downarrow},\ket{\out},\ket{\uparrow}\}$. However, preliminary measurements suggest that even in this situation, long-time dynamics are nearly exponential-like.   

On the theoretical side, it is a challenge to interpret and calculate the threshold power for Rabi oscillations. 
Furthermore, the qualitative similarities between the Rabi oscillations of the Fermi polaron and the much more complex spin-balanced unitary gas might point to intriguing opportunities to understand many-body systems without small parameters from a (seemingly unrelated) impurity limit~\cite{Meera_2016,levinsen2017}.

We thank S. Girvin and L. Glazman for fruitful discussions, and R. Fletcher, G. Martirosyan, M. Gemaljevic, and E. Papoutsis for comments on the manuscript. We thank M. Zwierlein and B. Mukherjee for sharing their data. This work was supported by the NSF (Grant No. PHY-1945324), AFOSR (Grant No. FA9550-23-1-0605), the David and Lucile Packard Foundation, and the Alfred P. Sloan Foundation. G.L.S. acknowledges support from the NSF Graduate Research Fellowship Program. A.S. acknowledges support from the U.S. Department of Energy, Office of Science, National Quantum Information Science Research Centers, Quantum Systems Accelerator.

\clearpage
\newpage

\setcounter{figure}{0}
\renewcommand{\thefigure}{S\arabic{figure}}
\setcounter{equation}{0}
\renewcommand{\theequation}{S\arabic{equation}}
\setcounter{secnumdepth}{3}

\title{Supplementary information\\Emergence of Fermi’s Golden Rule in the Probing of a Quantum Many-Body System}
\maketitle

\onecolumngrid
\renewcommand{\thefigure}{S\arabic{figure}}
\newcounter{mycounter}
\setcounter{mycounter}{1}
\newcommand{\RNum}[1]{\uppercase\expandafter{\romannumeral #1\relax.}}

\section{Density homogeneity of the initial and outcoupled internal states}
\stepcounter{mycounter}

\begin{figure}[!hbt]
\includegraphics[width=1\columnwidth]{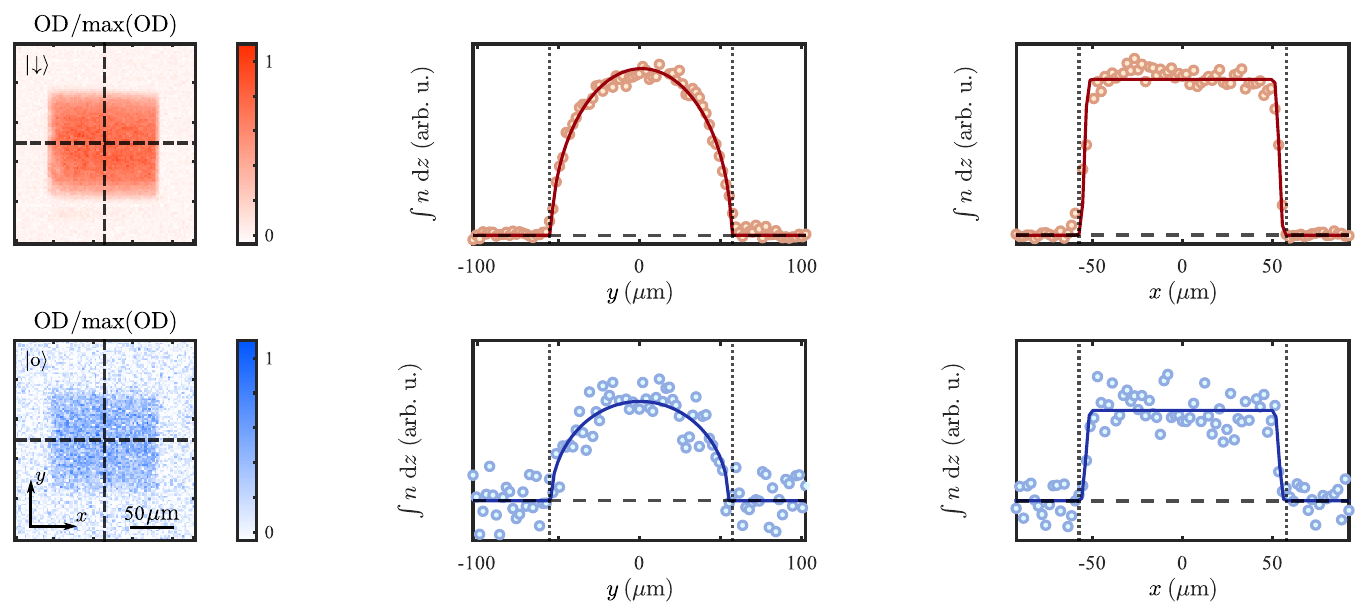}
\caption{Spatial homogeneity of the initial spin mixture and of the outcoupled state in the optical box. The upper (resp. lower) row corresponds to the state $\ket{\downarrow}$ before the coupling is turned on (resp. $\ket{\out}$ after the coupling pulse). The color code is the same as Fig.~\ref{Fig1} of the main text. The solid lines on the density profile cuts are fits assuming a uniform density distribution within a box trap. The vertical dotted lines mark the edges of the box trap and are located at the same position for each column. Here $N_\downarrow = 1.71(5)\times 10^5$ ($\approx N_\uparrow$) and $N_\out = 2.2(1) \times 10^4 $.}
\label{box configuration}
\end{figure}

The realization of uniform quantum gases confined in optical boxes greatly simplifies the interpretation of density-dependent phenomena and eliminates inhomogeneous broadening in spectroscopy~\cite{navon2021,vale_2021spectroscopic}. Here we characterize the density homogeneity of the initial spin mixture and of the outcoupled state. 
Our system initially consists of a two-component gas of $^6$Li fermions in a spin-balanced mixture. The atoms are trapped in a cylindrical optical box trap (for more technical details, see \emph{e.g.}~\cite{ji_2022,ji_2023_QJT}). Examples of \emph{in situ}\ absorption images of the occupied internal state $\ket{\downarrow}$ before the coupling is turned on and the final outcoupled state $\ket{\out}$ after the coupling pulse are shown on the left of Fig.~\ref{box configuration}. The density profiles correspond respectively to the cuts along the dashed line on the OD images (the middle and rightmost profiles are respectively the cuts along the radial and axial directions). These profiles agree very well with fits that assume a homogeneous density inside the box (see solid lines).

\section{Validity range of FGR away from the peak response frequency}
\begin{figure}[!hbt]
\includegraphics[width=0.57\columnwidth]{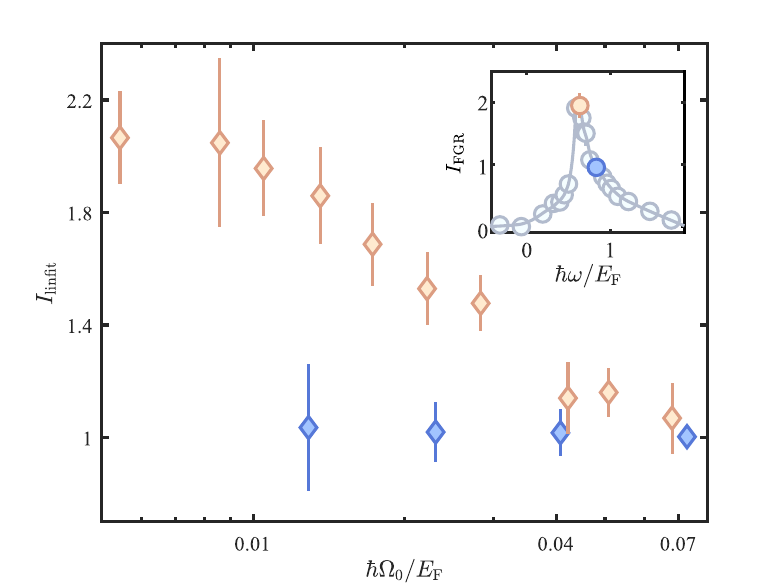}
\caption{Validity of FGR away from the resonance. The linear rate is extracted with a linear fit of the transfer fraction with pulse durations up to $t = 2\pi/(5\Omega_0)$.  
Inset: same spectrum as in the upper inset of Fig.~\ref{Fig3}, where color markers are adjusted to match the main panel, \emph{i.e.}\ the blue (resp. light orange) point corresponds to the detuned (resp. on-resonance) response. Note that for this spectrum, $\hbar\Gamma_0/E_\F \approx 0.24$.}
\label{off_peak_pd}
\end{figure}

\stepcounter{mycounter}
The validity range of FGR with respect to $\Omega_0$ is $\omega$ dependent and is most stringent on resonance, $\omega=\omega_{\mathrm{p}}$. In the main text, we observed the on-resonance power broadening of the spectral response. Here we briefly look into the applicability of FGR away from $\omega_{\mathrm{p}}$. 
In Fig.~\ref{off_peak_pd}, we show the measurement of the linear transfer rate $I_\text{linfit}$ for a detuned drive of frequency $\hbar\omega/E_{\mathrm{F}} = 0.76(2)$, which is approximately $\Gamma_0/2$ away from $\omega_{\text{p}}$ (see blue points in the inset of Fig.~\ref{off_peak_pd});  we use pulses of duration up to $2\pi/(5\Omega_0)$. We show this data alongside the original data of the lower panel of Fig.~\ref{Fig4} (light orange points). 
For the detuned drive, the response remains unchanged up to $\hbar\Omega_0/E_{\mathrm{F}} \approx 0.07$; in the resonant case, deviations can already be seen for $\hbar\Omega_0/E_{\mathrm{F}} \gtrsim 0.01$.   

\section{Extraction of spectral properties}
\stepcounter{mycounter}

\begin{figure}[!hbt]
\includegraphics[width=1\columnwidth]{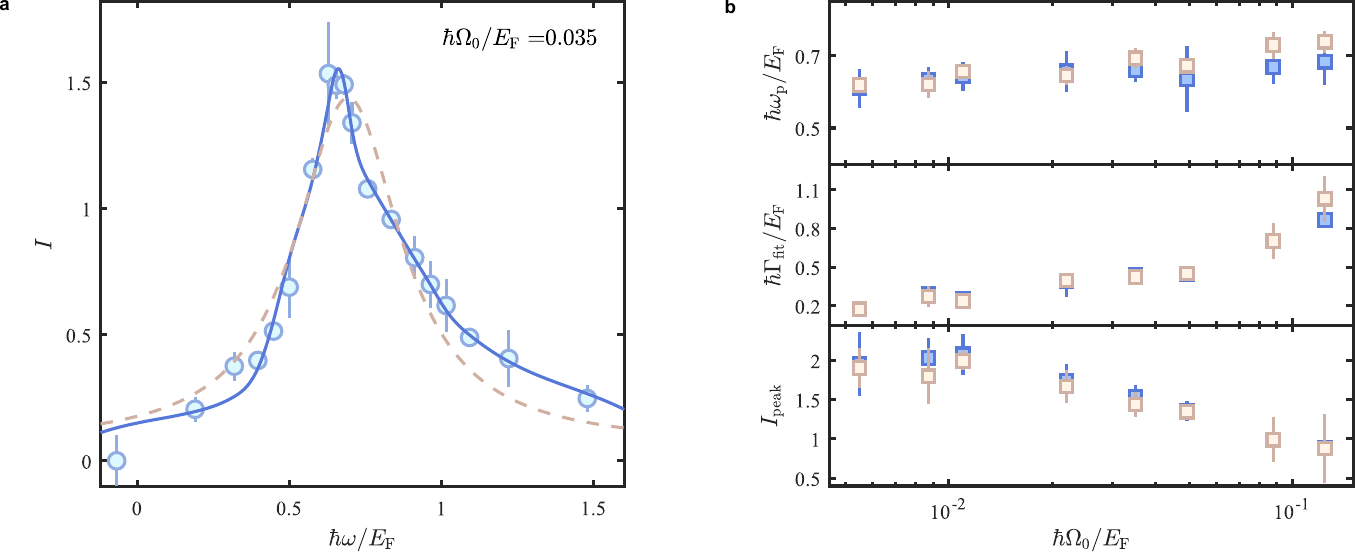}
\caption{Robustness of the extraction of the spectral property. (\textbf{a}) Example of an rf spectrum, together with two models for extracting the spectral properties. The solid blue (resp. dashed beige) line is a smoothened interpolation using an SG filter (resp. fitting with Lorentzian lineshape). (\textbf{b}) Comparison between the spectral properties extracted using the two models; the color style is the same as in (\textbf{a}).}
\label{spec_analysis}
\end{figure}
In Fig.~\ref{Fig4} of the main text, we extracted properties from the measured spectra using smoothened interpolations of the data via a Savitzky–Golay (SG) filter~\cite{SG_filter}. To verify the robustness of these extracted properties, we also use a Lorentzian fitting function. An example of a direct comparison on a typical spectrum is shown in Fig.~\ref{spec_analysis}a. In Fig.~\ref{spec_analysis}b, we also show the comparison of the extracted spectral properties. 
The parameters largely match (even though the Lorentzian fit performs badly in the tails of the spectrum).

\section{Calibration of $\Omega_0$ and Magnetic field stability}
\stepcounter{mycounter}

\begin{figure}[!hbt]
\includegraphics[width=1\columnwidth]{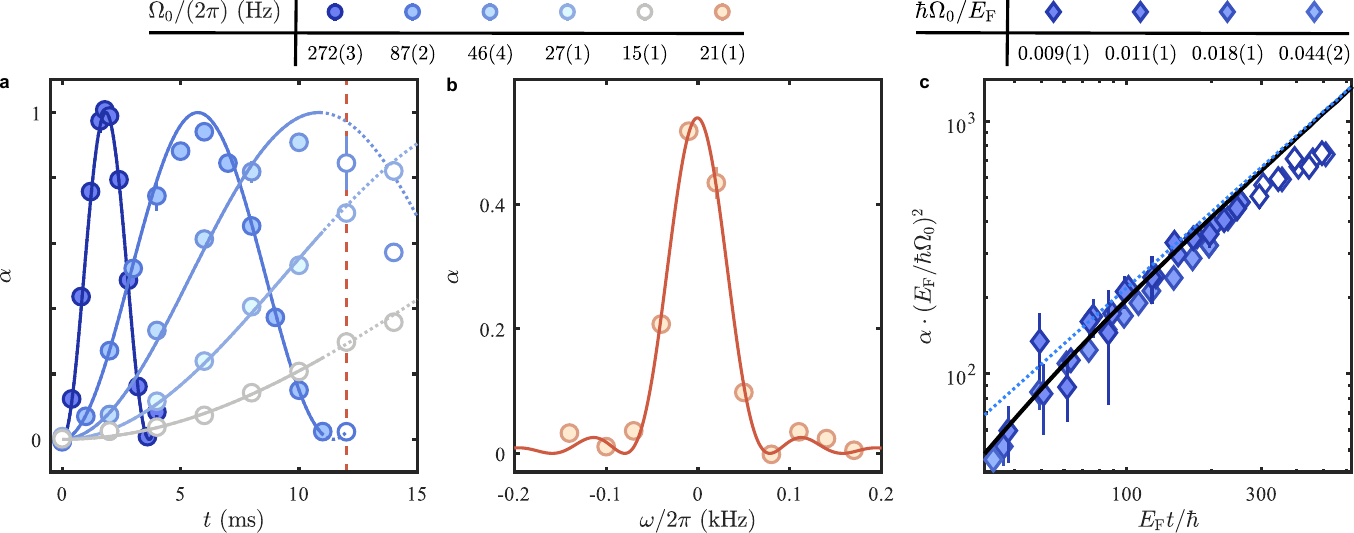}
\caption{Calibration of $\Omega_0$ and magnetic field stability. (\textbf{a}) Rabi oscillations in fully polarized gases using on-resonant rf. The solid lines are fits to the data. The dotted lines are the extension of the fits and the open markers are beyond the pulse duration where the Fourier-limited spectrum is achieved. (\textbf{b}) Rf spectrum of a fully polarized gas that is Fourier limited for a pulse duration of $t = 12~\mathrm{ms}$. The solid line is a sinc-squared fit. (\textbf{c}) Deviation from linear-response expectation due to magnetic field noise. The solid-marker data, the solid black line, and the dotted blue line are the same as in Fig.~\ref{Fig3}. Open markers are the measurements beyond our limit of $12~\mathrm{ms}$.
}
\label{Omega0_gallery}
\end{figure}

Here we present the calibration of the bare-transition Rabi frequency $\Omega_0$ between internal states $\ket{\downarrow}$ and $\ket{\out}$. We prepare a fully polarized Fermi gas in $\ket{\downarrow}$ and apply a resonant rf pulse for a duration $t$. Typical Rabi flops are shown in Fig.~\ref{Omega0_gallery}a, with the solid lines representing sine-squared fits to the data. 

To provide a bound to our magnetic field stability, we measure long-pulse spectra; in Fig.~\ref{Omega0_gallery}b, we show a spectrum taken with a pulse duration of $12~\mathrm{ms}$. The solid line is a fit to a sinc-squared function, whose FWHM is $2\pi\times 70(4)~\mathrm{Hz}$; this agrees well with the expected Fourier limit $\approx 2\pi \times 73.8~\mathrm{Hz}$. This implies that our magnetic field stability is better than $\approx 35~\mathrm{mG}$. 

As a result, we restricted in Fig.~\ref{Fig3} the data to pulse durations below $12~\mathrm{ms}$. In Figs.~\ref{Omega0_gallery}a and c, we show the data for pulse durations longer than $12~\mathrm{ms}$ as open markers. Although the transfer fraction is low ($\tsf<0.1$), we observe a deviation from the predicted solid black line. We attribute this to magnetic field drifts over longer times.

\section{Linear response theory: FGR and beyond}
\stepcounter{mycounter}

In this section, we theoretically outline the different regimes within the linear response region depicted in Fig.~\ref{Fig6}. 
The full Hamiltonian is
\begin{equation}\label{eq:fullH}    
\hat{H}(t)=\hat{H}_{\mathrm{S}}+\hat{V}_\mathrm{pr}(t)
\end{equation}
where $\hat{H}_{\mathrm{S}}$ is the (time-independent) Hamiltonian of the spin-$1/2$ unitary Fermi gas, with spin states $\ket{\uparrow}$ and $\ket{\downarrow}$. 
The probing term that couples states $\ket{\downarrow}$ and $\ket{\out}$ is
\begin{equation}
    \hat{V}_\mathrm{pr}(t)=\hbar\Omega_0 s(t)  \cos\left(\omega_\mathrm{pr} t\right) \sum_\kk \hat{c}_{\text{o},\kk}^\dagger \hat{c}_{\downarrow,\kk}+\mathrm{h.c.},
\end{equation}
where $\omega_{\mathrm{pr}}$ is the driving frequency, $\Omega_0$ is the bare-transition Rabi frequency, $s(t)$ is the temporal profile of the rf field (assumed to be a square pulse in the main text), and $\hat{c}_{\sigma,\kk}$ (resp. $\hat{c}_{\sigma,\kk}^\dagger$) is the annihilation (resp. creation) operator for particles with state $\sigma$ and momentum $\hbar\kk$. 
The transition rate into $\ket{\out}$ is our observable, defined as
\begin{equation}
\begin{aligned}
\hat{R}\equiv \frac{1}{N_\downarrow}\frac{\mathrm{d}\hat{N}_\out}{\mathrm{d}t} 
&=\frac{i}{\hbar N_\downarrow}[ \hat{V}_\mathrm{pr},\hat{N}_\out]\\
&=-i\frac{\Omega_0}{N_\downarrow}s(t)  \cos\left(\omega_\mathrm{pr} t\right) \sum_\kk \hat{c}_{\out,\kk}^\dagger \hat{c}_{\downarrow,\kk} +\mathrm{h.c.},
\end{aligned}
\label{trans_rate_def}
\end{equation}
where $\hat{N}_\out = \sum_\kk \hat{c}_{\out,\kk}^\dagger \hat{c}_{\out,\kk}$ is the number operator of the outcoupled particles, and $N_\downarrow$ is the initial number of particles in state $\ket{\downarrow}$.

\subsection{Linear response theory of the transfer rate}

We now calculate the time evolution of the transition rate as a function of drive detuning $\omega\equiv\omega_\mathrm{pr}-\omega_{0}$: 
\begin{equation}\label{eqn: spin current linear response}
\begin{aligned}
R(\omega,t)\equiv \langle\hat{R}\rangle&\approx- \frac{i}{\hbar }\int_{-\infty}^\infty \mathrm{d}t'\Theta(t-t')\Big\langle[\hat{R}^\mathrm{I}(t),\hat{V}_\mathrm{pr}^\mathrm{I}(t')]\Big\rangle_0\\
& \approx\frac{\Omega_0^2}{4 N_\downarrow}\sum_\kk\int_0^t\dd t'~e^{i(\xi_{k\downarrow}/\hbar-\omega)(t-t')}(-i)G_{\downarrow}^<(\kk,t-t')+\mathrm{c.c.}\\
&=\frac{\Omega_0^2}{4 N_\downarrow}\sum_\kk\int_0^t\dd t' \int_{-\infty}^\infty\frac{\dd\omega'}{2\pi}~e^{i(\xi_{k\downarrow}/\hbar-\omega-\omega')(t-t')}A(\kk,\omega')n_{\mathrm{F}}(\hbar\omega')+\mathrm{c.c.},
\end{aligned}
\end{equation}
where $\hat{\mathcal{O}}^\mathrm{I} \equiv e^{i\hat{H}_{\mathrm{S}} t/\hbar}\hat{\mathcal{O}}e^{-i\hat{H}_{\mathrm{S}}t/\hbar}$ is the corresponding operator in the interaction picture. The approximations in the first and second lines are respectively linear response (\emph{i.e.} we only keep the term to first order in $\Omega_0^2$) and rotating wave approximations.   
$\braket{\hat{\mathcal{O}}}_0\equiv\mathrm{Tr}[\hat{\rho}_0\hat{\mathcal{O}}]$ denotes the average over the initial equilibrium state $\hat{\rho}_0$ of a spin-balanced spin-1/2 unitary gas governed by the Hamiltonian $\hat{H}_{\mathrm{S}}$. $\braket{\hat{\mathcal{O}}}\equiv \mathrm{Tr}[\hat{\rho}(t)\hat{\mathcal{O}}]$ denotes the average over the state $\hat{\rho}(t)$ which at $t=0$ is the state $\hat{\rho}_0$ and which evolves according to the full time-dependent Hamiltonian Eq.~\eqref{eq:fullH}. We introduced the lesser Green's function $G_{\downarrow}^<(\kk,t-t')\equiv i e^{i\mu_\downarrow(t-t')}\braket{ (\hat{c}_{\downarrow,\kk}^\mathrm{I}(t') )^{\dagger}\hat{c}_{\downarrow,\kk}^\mathrm{I}(t)}_0$, and the single-particle kinetic energy  $\xi_{k\downarrow} =\hbar^2k^2/(2m)-\mu_\downarrow$ with $k = \lvert \kk \rvert$ and chemical potential $\mu_\downarrow$. The spectral function $A(\kk,\omega)$ is related to the Fourier transform of the lesser Green's function through $A(\kk,\omega)n_{\mathrm{F}}(\hbar\omega) = G_{\downarrow}^<(\kk,\omega) \equiv \int_{-\infty}^{\infty}e^{i\omega t}G_{\downarrow}^<(\kk,t) \dd t $, where $n_\F(\epsilon) = 1/(e^{{\beta}\epsilon}+1)$ is the Fermi-Dirac distribution, and $\beta^{-1}\equiv k_\mathrm{B}T$. In the second line, we assumed that the coupling is abruptly turned on at $t=0$.  

\subsection{Long-time limit: FGR}

In the long-time limit (while remaining within the validity regime of linear response theory), Eq.~\eqref{eqn: spin current linear response} reduces to FGR, which provides a simple expression for the rf spectrum $R_{\mathrm{FGR}}(\omega)$:
\begin{equation}\label{eqn: I long time}
    R_\mathrm{FGR}(\omega)\equiv \frac{\Omega_0^2}{4N_{\downarrow}}\sum_\kk A(\kk,\xi_{k\downarrow}/\hbar-\omega)n_{\mathrm{F}}(\xi_{k\downarrow}-\hbar\omega).
\end{equation}
The normalization of $R_\mathrm{FGR}(\omega)$ is 
\begin{equation}\label{eqn: norm Irf}
    \int_{-\infty}^\infty\frac{\dd\omega}{2\pi}~R_\mathrm{FGR}(\omega)=\frac{\Omega_0^2}{4}.
\end{equation}
In the main text, we defined a normalized spectrum  $I_\mathrm{FGR}(\hbar\omega/E_{\mathrm{F}}) \equiv R_\mathrm{FGR}(\omega)\cdot(E_{\mathrm{F}}/\hbar\Omega_0^2)$. This spectrum is normalized such that 
\begin{equation}\label{eqn: norm Irf}
    \int_{-\infty}^\infty \dd \left(\frac{\hbar\omega}{E_{\mathrm{F}}}\right)~I_\mathrm{FGR}\left(\frac{\hbar\omega}{E_{\mathrm{F}}}\right)=\frac{\pi}{2}.
\end{equation}

\subsection{Relation between linear response and FGR}
The time evolution Eq.~\eqref{eqn: spin current linear response} can be related to the long-time response Eq.~\eqref{eqn: I long time} by a change of variable $\omega'\rightarrow\xi_{k\downarrow}/\hbar-\omega'$
\begin{equation}\label{eqn: relation I ejection short and long time}
    \begin{aligned}
        R(\omega,t)&=2\cdot\mathrm{Re}\left[\int_0^t\dd t'\int_{-\infty}^\infty \frac{\dd\omega'}{2\pi} ~ e^{i(\omega'-\omega)t'}R_\mathrm{FGR}(\omega')\right]\\
        &=2t\int_{-\infty}^\infty\frac{\dd \omega'}{2\pi}~\mathrm{sinc}\Big[(\omega'-\omega)t\Big]R_\mathrm{FGR}(\omega').
    \end{aligned}
\end{equation}
We obtain Eq.~(\ref{relation_N_I}) from the main text via integration
\begin{equation}\label{eqn: tsf square}
    \tsf(\omega,t)=\int_0^t \dd t'~R(\omega,t') = t^2\int_{-\infty}^\infty\frac{\dd\omega'}{2\pi}~\mathrm{sinc}^2\left(\frac{\omega'-\omega}{2}t\right)R_\mathrm{FGR}(\omega').
\end{equation}

\subsection{Short-time limit: quadratic dependence in time}
At fixed $\omega$, the leading-order transition rate Eq.~\eqref{eqn: relation I ejection short and long time} for small $t$ is
\begin{equation}
    R(\omega,t)\approx 2t\int_{-\infty}^\infty\frac{\dd\omega'}{2\pi}R_\mathrm{FGR}(\omega')=\frac{\Omega_0^2 t}{2},
\end{equation}
where we have used the normalization condition Eq.~\eqref{eqn: norm Irf}. The transfer fraction is then
\begin{equation}
    \tsf \approx \frac{\Omega_0^2t^2}{4}.
\end{equation}
Note that the sub-leading order can be calculated using the contact $C$, which characterizes the high-frequency limit of the rf spectrum~\cite{Tan_2008,schneider_2010,braaten_2011}:
\begin{equation}
    R_\mathrm{FGR}\left(\frac{\hbar\omega}{\EF} \gg 1\right)\approx\frac{\sqrt{\hbar}\Omega_0^2}{8\pi N_{\downarrow}}\frac{C}{\sqrt{m}\omega^{3/2}}.
\end{equation}
The sub-leading correction to the transfer fraction is thus $\propto t^{5/2}$:
\begin{equation}
    \tsf(\omega,t)\approx \frac{\Omega_0^2}{4}t^2\left[1-\frac{4}{15}\frac{\tilde{C}}{\pi^{3/2}}\left(\frac{t}{t_\F}\right)^{1/2}\left(1-\frac{3}{7}\omega t\right)+\cdots\right],
\end{equation}
where $t_\F\equiv \hbar/E_\F$, and $\tilde{C}\equiv C/(N_{\downarrow}k_\F)$.

\subsection{Generalization for arbitrary pulse form}
In most of this work, we assumed square coupling pulses. We now consider more general pulse functions $s(t)$ as defined in Eq.~(\ref{trans_rate_def}), and find a generalized form of Eq.~(\ref{eqn: tsf square}): \begin{equation}
    \alpha(\omega,t)=\int_{-\infty}^\infty\frac{\dd \omega'}{2\pi}\left|\int_{0}^t\dd t' ~s(t') e^{i(\omega'-\omega)t'}\right|^2 R_\text{FGR}(\omega').
\end{equation}
The transfer fraction in the linear response regime is a convolution of the truncated Fourier transform of $s(t)$ with $R_\text{FGR}$. 

The long-time FGR limit is
\begin{equation}
    \alpha(\omega,\EF t\gg 1)=\mathcal{A}_\text{rf}R_\text{FGR}(\omega),
\end{equation}
where $\mathcal{A}_\text{rf}\equiv\int_{-\infty}^\infty\dd t|s(t)|^2$ is the pulse area.

\section{Short-time extrapolations for Fig.~\ref{Fig3}}
\stepcounter{mycounter}
The short-time behavior of Eq.~(\ref{relation_N_I}) is governed by the high-frequency part of the rf spectrum. In this section, we extrapolate the theoretical prediction shown in the inset of Fig.~\ref{Fig3} to short times. The black dashed line corresponds to $\tilde{C} = 2.85$ (at $T/T_{\mathrm{F}} \approx 0.15$)~\cite{Mukherjee_2019,Carcy_2019}; the orange shaded area represents an error band corresponding to different high-frequency extrapolations (see caption).

\begin{figure}[!hbt]
\includegraphics[width=1\columnwidth]{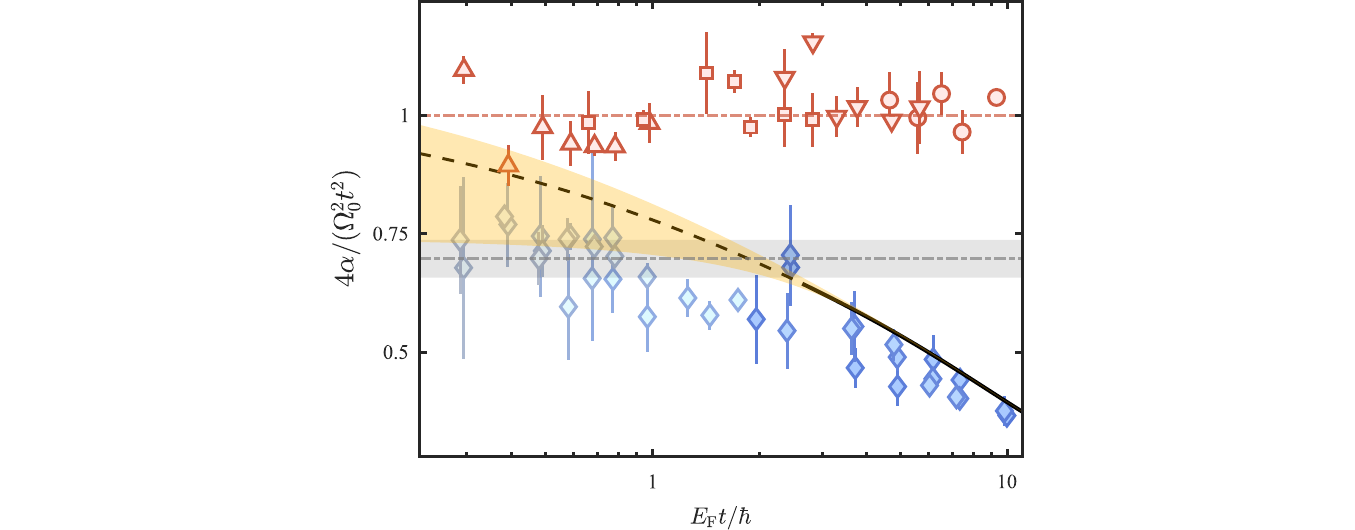}
\caption{Short-time extrapolations for the theoretical prediction for $\tsf(t)$. The markers and curves are the same as in the inset of Fig.~\ref{Fig3}. The black dashed line corresponds to $\tilde{C} = 2.85$. The orange error band is set on the low end by truncating the integral in Eq.~(\ref{relation_N_I}) to the measurement frequency range, and on the high end, by completing the data with a tail of contact $\tilde{C}=3.5$ (a reasonable upper value set by recent measurements~\cite{Mukherjee_2019,Carcy_2019}).
}
\label{shortt_exp}
\end{figure}

\section{Rabi oscillation properties in the unpolarized and highly-polarized limits} 
\stepcounter{mycounter}
In the main text, we noticed that both the low-$\Omega_0$ limit for $(\Omega_{\mathrm{R}}/\Omega_0)^2$ and the threshold $\Omega_\text{thres}/\Gamma_0$ for the unpolarized (\emph{i.e.}\ spin-balanced) measurements are close to those measurements in the highly-polarized (Fermi polaron) limit~\cite{Vivanco_2023}. This is especially intriguing as $(\Omega_{\mathrm{R}}/\Omega_0)^2$ coincides with the Fermi polaron quasiparticle residue in the low-$\Omega_0$ limit, while such an interpretation is not straightforward in the unpolarized gas. In this section, we present a systematic comparison of the Rabi oscillation properties between the unpolarized and highly-polarized cases.

\begin{figure}[!hbt]
\includegraphics[width= 1\columnwidth]{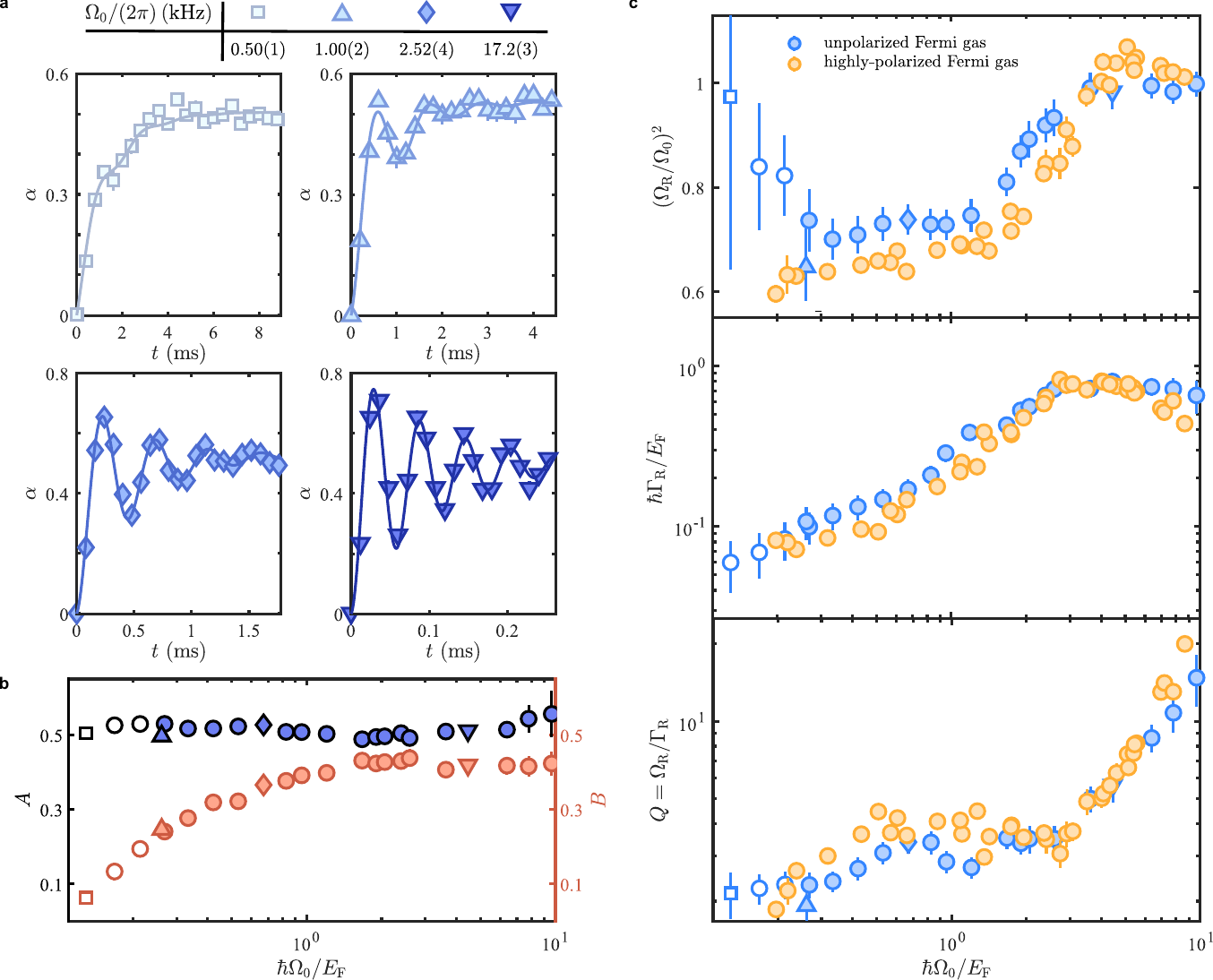}
\caption{Rabi oscillation spectroscopy for the spin-balanced and highly polarized Fermi gases. (\textbf{a}) A gallery of time series for the unpolarized gas for various $\Omega_0$. The solid lines are fits to the function $\tsf(t) = A\left(1-e^{-\Gamma_{\mathrm{R}}t/2}\right) - B\left[\mathrm{cos}(\Omega_{\mathrm{R}}t)-1\right]e^{-\Gamma_{\mathrm{R}}t/2}$. (\textbf{b}) The corresponding fitting parameters $A$ and $B$ (see text). (\textbf{c}) Comparison of the extracted parameters for the unpolarized (blue points) and the highly-polarized (orange points)~\cite{Vivanco_2023}. The same marker style is used in (\textbf{a}), (\textbf{b}) and (\textbf{c}).}
\label{gall_comp}
\end{figure}

In Fig.~\ref{gall_comp}a, we show four examples of the time series data. The first (squares) and second series (upper triangles) are respectively below and above the threshold Rabi frequency (recall that the Fermi energy is $E_{\mathrm{F}}/h \approx 4~$kHz). The solid lines are fits to a simple form $\tsf(t) = A\left(1-e^{-\Gamma_{\mathrm{R}}t/2}\right) - B\left[\mathrm{cos}(\Omega_{\mathrm{R}}t)-1\right]e^{-\Gamma_{\mathrm{R}}t/2}$; the parameters $A$ and $B$ are shown in Fig.~\ref{gall_comp}b. The value of $A$ gives the long-time asymptotic value of $\tsf$, and it remains $\approx0.5$ across a wide range of $\Omega_0$. The value of $B$ gives the oscillation amplitude; it increases monotonically with $\Omega_0$, indicating higher-quality oscillations at large $\Omega_0$. In Fig.~\ref{gall_comp}c, we show the renormalized oscillation frequency $\Omega_\mathrm{R}$, the decay rate $\Gamma_\mathrm{R}$ and their ratio (the quality factor $Q$). It is striking that for the data sets are quantitatively very close. 

In the absence of well-defined quasiparticles in the normal unpolarized unitary gas, it is surprising that $\Gamma_{\mathrm{R}}$ is comparable to that of the highly-polarized Fermi gas (which is a good Fermi liquid of polarons~\cite{yan_2019}), as shown by the middle panel of Fig.~\ref{gall_comp}c. Even the positions of the local maximum around $\Omega_0\approx 4E_\mathrm{F}/\hbar$ are similar.

These similarities are particularly remarkable since the low-$\Omega_0$ plateau of $\Omega_\mathrm{R}$ and the local maximum of $\Gamma_\mathrm{R}$ have specific interpretations in terms of Fermi-polaron physics (respectively, the quasiparticle residue and the hybridization of a dressed polaron state with the broad-continuum remnant of the repulsive unitary polaron~\cite{Vivanco_2023}) without obvious counterparts for the unpolarized gas. This suggests that the impurity limit of the Fermi polaron could provide a useful framework to understand the seemingly very different unpolarized unitary gas.

\end{document}